\documentclass[apjl]{emulateapj}
\usepackage{amsmath}
\def\dim#1{\mbox{\,#1}}
\newcommand \chimps{\ h^{-1} \dim{Mpc}}

\begin{document}

  \title{AGN Outflows and the Matter Power Spectrum}
  
  \author{Robyn Levine\altaffilmark{1,2}}
  \author{Nickolay Y. Gnedin\altaffilmark{2,3}}
  \altaffiltext{1}{JILA, University of Colorado, Boulder, CO 80309; robyn.levine@colorado.edu}
  \altaffiltext{2}{Particle Astrophysics Center, Fermi National Accelerator Laboratory, Batavia, IL 60510; gnedin@fnal.gov}
  \altaffiltext{3}{Department of Astronomy \& Astrophysics, The
  University of Chicago, Chicago, IL 60637}

  
  \begin{abstract}
    We have investigated the effects of AGN outflows on the amplitude
    of the matter power spectrum in a simple model of spherically
    symmetric outflows around realistically clustered AGN
    population. We find that two competing effects influence the
    matter power spectrum in two opposite directions.  First, AGN
    outflows move baryons from high to low density regions, decreasing
    the amplitude of the matter power spectrum by up to 20\%. Second,
    high clustering of the AGN transfers the power from small to
    larger scales. The exact balance between these two effects depends
    on the details of outflows on small scales, and quantitative
    estimates will require much more sophisticated modeling than
    presented here. 
  \end{abstract}
  
  \keywords{cosmological parameters---cosmology: theory---galaxies:
  active---intergalactic medium---large-scale structure of universe}


  \section{INTRODUCTION}

  Cosmic shear from weak lensing depends on the matter distribution on
  all scales, whether the matter is composed of baryons or dark
  matter.  Because it depends on the mass distribution alone, it is
  also independent of the dynamics of the matter.  Weak lensing can
  offer constraints on the equation of state of dark energy, as it is
  influenced by the expansion history of the universe.  In order to
  use weak lensing to distinguish between different dark energy
  models, constraints on cosmological parameters determined from high
  precision measurements of the matter power spectrum require accurate
  modeling of the physics on the relevant scales
  \citep[e.g.][]{DESWhite05}.  \citet{Haganetal05} have shown the
  importance of resolving dark matter substructure for simulating the
  nonlinear power spectrum.  Baryons are assumed to closely follow the
  underlying dark matter distribution on large scales, however, on
  smaller scales baryons are taking part in more complicated physical
  processes which can result in a signal comparable to, if not larger
  than the statistical errors on the matter power spectrum
  \citep{White04, ZhanKnox04, Jingetal06, Ruddetal06}.  In the recent
  release of the 3rd year data from the {\it Wilkinson Microwave
  Anisotropy Probe} (WMAP) a comparison of cosmic microwave background
  (CMB) measurements of the amplitude of the matter power spectrum
  with those of other methods finds weak lensing results to be the
  most discrepant \citep{Spergeletal06}. This discrepancy may simply
  be a statistical fluctuation, but might also reflect the complicated
  role the baryons can play in shaping the matter power spectrum at
  10\% level.

  The colossal energy input from active galactic nuclei (AGNs) may
  also affect the clustering of matter.  AGNs are known to influence
  their environments out to Mpc scales, creating bubbles of hot,
  tenuous gas around their host galaxies.  With kinetic energies
  corresponding to as little as a percent of their bolometric
  luminosities, AGNs produce outflows energetic enough to fill large
  fractions of the intergalactic medium (IGM) with very low density
  bubbles \citep[][hereafter LG05]{LevineGnedin05}.  These bubbles
  effectively push gas aside, influencing the distribution of baryons
  out to large scales. If AGN outflows are energetic enough to
  influence the large scale distribution of baryons, it is possible
  that descriptions of the matter distribution on these scales will
  need to include them.

  In this paper, we investigate the effect of AGN outflows on the
  matter power spectrum.  We
  continue with the simple model of LG05 and compare the influence of
  outflows of different kinetic energies on the matter distribution.
  In Section \ref{sec:sim} we describe the outflow model, and our
  model of the matter distribution and the corresponding power
  spectrum.  In Section \ref{sec:res}, we show the effects on the
  power spectrum for different simulation box sizes, resolutions, and
  outflow energies.  Section \ref{sec:con} is a summary of our
  findings and their implications.


  \section{SIMULATION}
  \label{sec:sim}

  In the following subsections, we briefly overview a simple model of
  AGN outflows and the effect they might have on the cosmic density
  distribution.  LG05 describes in greater detail the assumptions
  behind the outflow model.
  

  \subsection{A Simple Outflow Model}
  \label{subsection:out}

  Using a particle-mesh code, we simulate an evolving dark matter
  distribution under the assumption that the gas distribution follows
  that of the dark matter.  We determine the distribution of AGNs
  within the simulation by introducing a simple constant bias that
  assumes AGNs will lie in high density regions.  By combining a quasar
  luminosity function constrained by observations
  \citep{Cristiani04,Fanetal01a,Fanetal01b,SB03,Boyle00} with some
  simple arguments about the fraction of AGNs expected to host outflows
  at any given epoch, we obtain the number of sources to include in
  the simulation.

  We assume that following a brief energy injection from the AGN,
  spherically symmetric outflows expand according to the Sedov-Taylor
  blast wave model until reaching pressure equilibrium with their
  environments.  They then remain in pressure equilibrium, and any
  subsequent expansion (or contraction) is due to the Hubble expansion and the
  evolution of the cosmic density distribution within their
  neighborhoods.  The kinetic energy of the outflow is assumed to be a
  fixed fraction, $\varepsilon_k$ of the AGN's bolometric luminosity.
  Part of this study is to evaluate the role of different kinetic
  fractions on the large-scale matter distribution.

  In the above outflow model, AGNs fill large spherical bubbles with
  hot, tenuous gas, pushing aside the gas in the IGM in the process.
  In this model, the intergalactic gas is compressed into thin shells
  around the outflows.  In the calculation of the power spectrum in
  the next section, we determine the total matter density distribution
  by including these thin shells of dense matter surrounding the
  outflows into our model.


  \subsection{Determining the Matter Power Spectrum}
  \label{subsection:pow}

  We define the matter density in each cell of the simulation box
  according to:

  \begin{align*}
    1+\delta_{m} = \frac{\Omega_{dm}}{\Omega_m}(1+\delta_{dm}) +
    \frac{f_b\ \Omega_b}{\Omega_m}(1+\delta_{dm})
  \end{align*}
  \begin{align} \label{eq:dtot}
     = (1+\delta_{dm}) \frac{(\Omega_{dm} + f_b\ \Omega_b)}{\Omega_m},
  \end{align}
    
  \noindent where $f_b$ is a parameter that determines the local
  baryon fraction.  We assume that $\Omega_{dm} = \Omega_m -
  \Omega_b$, with $0.27$ and $0.04$ for $\Omega_m$ and $\Omega_b$,
  respectively.  Using the dark matter and outflow distributions, we
  calculate the total matter density in the entire simulation volume.
  We assume that cells lying within outflow regions are basically
  devoid of baryons, and we take $f_b = 0$ in those cells.  In cells
  untouched by outflows, we assume the baryon distribution directly
  traces that of the dark matter and we take $f_b = 1$.  At the
  outflow boundaries, the weighting parameter is constrained by the
  average baryon density.  We divide the boundary cells into two
  types: inner and outer boundaries.  The weighting parameter for the
  inner boundary is manually set to lie between $0$ and $1$, and the
  weighting parameter for the outer boundary is then determined by the
  normalization.  The results do not appear to have a significant
  dependence on our choice of $f_b$ at the bubble boundaries.  The
  values adopted for $f_b$ in the present model are an
  over-simplification, and will provide an upper-limit for the effects
  of AGN outflows on the matter power spectrum.

  We calculate the matter power spectrum for the case in which baryons
  closely follow the dark matter distribution (without AGN outflows)
  and for the case in which outflows redistribute baryons.
  Specifically we are interested in the quantity
  \begin{equation}
    \Delta P(k)_{\rm AGN} = \frac{P(k)_{\rm AGN}}{P(k)_{\rm nobar}} - 1,
  \end{equation}
  where $P(k)_{\rm AGN}$ is the power spectrum including the
  redistribution of baryons by AGNs and $P(k)_{\rm nobar}$ is the power
  spectrum determined from dark matter alone (with baryons tracing
  dark matter on the scales we resolve).

  The outflows affect the large scale power via two competing effects
  in the above model.  First, outflows move gas around, redistributing
  baryons from high to low density regions.  This has the effect of
  decreasing the matter power on a large range of
  scales.  Second, if AGN outflows only affect the small-scale, immediate
  environment of host galaxies, high clustering of AGNs transfers the
  fluctuations from small to large scales.  This can be thought of
  symbolically by representing the gas density as
  \begin{equation}\label{eq:rhog}
    \rho_g(\mathbf{x}) = \hat{LS}[\rho_{g,0}(\mathbf{x})] + \hat{SS}[n_{AGN}(\mathbf{x})],
  \end{equation}
  where the gas density $\rho_g(\mathbf{x})$ is represented
  as the sum of the Large-Scale (LS) redistribution of the
  ``undisturbed'' gas density $\rho_{g,0}$ and additional Small-Scale
  (SS) density fluctuations caused by complex gas dynamical motions
  around AGNs. The quantities $\hat{LS}$ and $\hat{SS}$ are, in fact,
  operators in the strict mathematical sense, but if we approximate
  them as convolutions with some window functions, then the baryonic
  power spectrum can be symbolically represented as
  \begin{equation}\label{eq:pkg}
    P_{g}(k) = W^2_{LS}P_{g,0}(k) + W^2_{SS}P_{AGN}(k) + {\rm Cross\ Terms},
  \end{equation}
  where the Large-Scale factor $W^2_{LS}$ is, generally smaller than
  1, since one would expect the AGN outflows to move gas from high to
  low densities, thus reducing the clustering of gas (although, of
  course, the AGN effect may be very small, in which case $W^2_{LS}$
  could be indistinguishable from unity). The Small-Scale factor
  $W^2_{SS}$ is likely to be small, since only a small fraction of all
  galaxies host AGNs at any given moment, but it is multiplied by a
  large factor $P_{AGN}(k)$ - the latter is large since AGNs are highly
  clustered, with bias factors easily as high as 5, depending on
  redshift and luminosity \citep[e.g.][]{Croometal05, Lidzetal06}. In our
  model, $W^2_{SS}$ represents spherically symmetric bubbles, which
  behave similarly for all AGNs in the simulation.  In reality, it
  should be a much more complex factor, depending on the relevant
  physics and the AGN environment on smaller scales.  The contribution
  of the second term in Equation \ref{eq:pkg} also depends on the
  amount of bias in the AGN distribution.  A high bias factor can
  increase the size of the AGN term significantly.  The important
  point to take is that the AGN term need not be very large (10\%) to
  change the amplitude of the power spectrum by an amount sufficient
  to bias lensing measurements (1\%).


  \section{RESULTS}
  \label{sec:res}

  In Sections \ref{subsection:conv} and \ref{subsection:kinfrac}, we
  conduct convergence studies of simulation box size and resolution
  and we study the effects of different energy inputs from the AGNs on
  the amplitude of the matter power spectrum by varying the kinetic
  fraction, $\varepsilon_k$.  

  Figure 7 of \citet{Spergeletal06} shows a comparison of CMB
  predictions for the cosmological parameters $\Omega_m$ and $\sigma_8$
  with the weak lensing predictions from the first analysis of the
  Canada-France-Hawaii Telescope Legacy Survey (CFHTLS)
  \citep{Hoekstraetal05}.  We show the following power spectrum
  results for $z=0.81$, corresponding to the mean redshift of sources
  used in the CFHTLS, so that we might comment on their relevance to
  discrepancies between the two methods.

  \subsection{Convergence Studies}
  \label{subsection:conv}

\begin{figure}[t] 
    \centering
    \epsscale{1.2}
    \plotone{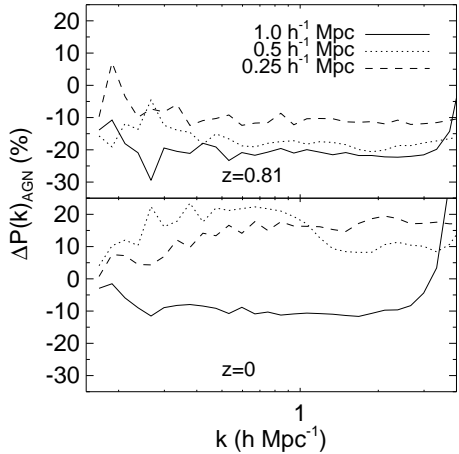}
    \caption{\label{fig:pres} Percentage difference between the matter
    power spectrum with and without the AGN outflow model for
    simulations of differing resolutions at two different
    redshifts. Each box is $64 h^{-1}\dim{Mpc}$ across.}
  \end{figure}

  To test the effects of resolution on the matter power spectrum
  results, we have calculated the power spectrum for simulation boxes
  of length $64 \chimps$ with resolutions of $1$, $0.5$, and $0.25
  \chimps$.  Figure \ref{fig:pres} shows the resulting $\Delta
  P(k)_{\rm AGN}$ at two different redshifts.  At higher redshift, the
  results are somewhat similar for simulations of different
  resolutions, showing a negative contribution to the amplitude of the
  matter power spectrum.  At $z=0$, the two higher resolution runs
  both show positive contributions to the power spectrum.  In the
  high-$z$ case, the AGN outflows move gas around, effectively
  lowering the amplitude of the power spectrum, while at $z=0$ it is
  possible that the effects of AGN clustering dominate, at least for
  the higher resolution runs.

\begin{figure}[t] 
    \centering
    \epsscale{1.0}
    \plotone{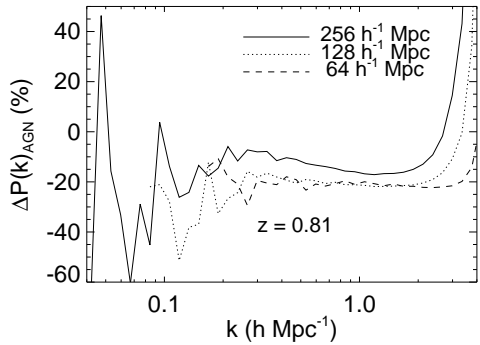}
    \caption{\label{fig:pbox} Percentage difference between the matter
    power spectrum with and without the AGN outflow model for
    different simulations volumes. Each box has cells of length $1
    h^{-1}\dim{Mpc}$.}
  \end{figure}

  Figure \ref{fig:pbox} shows a comparison of the results for boxes of
  different lengths, but each with the same resolution of $1 \chimps$.
  The box lengths shown are $64$, $128$, and $256 \chimps$.  The
  results for each box size are fairly similar, each showing a
  negative contribution of $\sim 20\%$ to the amplitude of the power
  spectrum. 

\begin{figure}[t]
    \centering
    \epsscale{1.0}
    \plotone{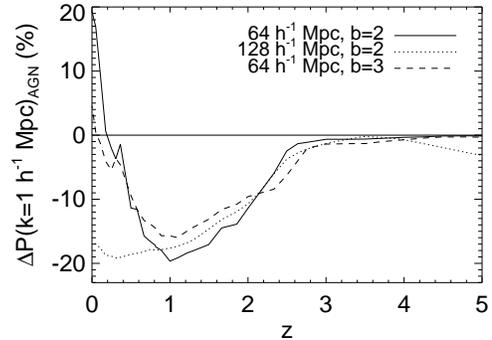}
    \caption{\label{fig:pz} Redshift evolution of the percentage
    difference between the power with and without the AGN outflow
    model at $k=1\ h \dim{Mpc}^{-1}$ for $0.5 \chimps$ resolution
    boxes.  Two different box sizes and two different bias factors are
    shown.  The turnaround in $\Delta P(k)_{\rm AGN}$ could be a
    result of AGN clustering effects dominating over the
    redistribution of gas by outflows.}
  \end{figure}

  Figure \ref{fig:pz} shows the redshift evolution of $\Delta
  P(k)_{\rm AGN}$ for boxes of length $64$ and $128 \chimps$, with
  $0.5 \chimps$ resolution, as well as a $64 \chimps$ box with a
  higher bias factor $b$.  For $z>1$, $\Delta P(k)_{\rm AGN}$
  decreases as the outflows expand, moving gas onto larger scales.  As
  the the filling fraction of AGN outflows levels off at low
  redshifts, the positive contribution of AGN clustering begins to
  take over, causing the turnaround in $\Delta P(k)_{\rm AGN}$ shown
  for the smaller volume box.  At lower redshifts, the larger volume
  simulation contains more bright AGNs (larger bubbles) than the
  smaller volume simulation, so the negative contribution of AGN
  outflows to the power spectrum continues to dominate.  The run with
  the higher bias factor also demonstrates the positive contribution
  of AGN clustering to $\Delta P(k)_{\rm AGN}$, since $\Delta
  P(k)_{\rm AGN}$ is greater for $b=3$ than for $b=2$.

  The convergence studies support our interpretation of the
  Large-Scale redistribution term in Equation \ref{eq:pkg}, because in
  the case where the negative contribution of outflows dominates (in
  the low-resolution simulations; see Figure \ref{fig:pbox}), some
  convergence takes place.  In the resolution study, as demonstrated
  by Figures \ref{fig:pres} and \ref{fig:pz}, convergence is not as
  obvious, as we do not model the small-scale fluctuations (the second
  term in Equation\ref{eq:pkg}) in detail.


  \subsection{Dependence of the Matter Power Spectrum on Kinetic Fraction}
  \label{subsection:kinfrac}

\begin{figure}[t] 
    \centering
    \epsscale{1.0}
    \plotone{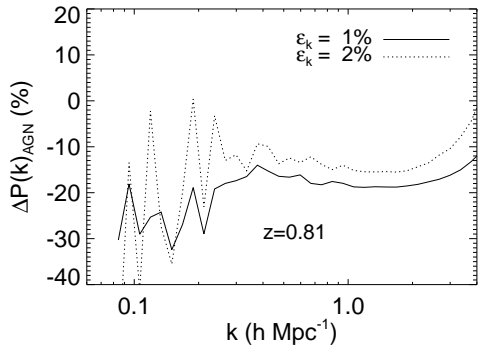}
    \caption{\label{fig:pkin} Percentage difference between the matter
    power spectrum with and without the AGN outflow model for models
    with different kinetic fractions, $\varepsilon_k$.}
  \end{figure}

  The kinetic energy driving outflows in AGNs is likely linked to the
  luminosity of the AGNs, but the exact fraction, $\varepsilon_k$ is
  not yet well constrained by observations.  As studies of the filling
  fraction of AGN outflows showed in LG05, the more kinetic energy
  input from the quasar, the greater the effect on the AGN
  environments.  In this simple outflow model, it only takes a very
  small fraction of the energy output of AGNs to produce outflows that
  fill the entire IGM by $z=2$.  In our convergence studies, we have
  adopted a kinetic fraction of $1\%$.  Here we examine the effects of
  varying the kinetic fraction on the power spectrum.  Figure
  \ref{fig:pkin} shows that as little as $1\%$ of an AGN's bolometric
  energy output produces changes to the power spectrum of up to $20\%$
  in the simple model presented here.  A kinetic fraction of $2\%$, or
  a doubling in the efficiency of the AGN, does not drastically change
  the effects on the matter power spectrum.  The precise dependence of
  the amplitude of the power spectrum on the kinetic fraction likely
  depends on the properties of AGNs in a non-trivial way.

  \section{DISCUSSION AND CONCLUSIONS}
  \label{sec:con}
  
  We have found that a simple model in which AGN outflows play a role
  in the distribution of baryonic matter on cosmic scales results in
  more than several percent difference in the amplitude of the matter
  power spectrum. Two competing effects - the removal of gas from high
  density regions and high clustering of AGNs - make contributions of
  opposite signs to the matter power.

  The amount of energy released in observed AGNs may be sufficient to
  move all of baryons in the universe over cosmological distances,
  which would result in the reduction in the matter power spectrum by
  up to 30\%.  Observationally, we do not really know whether AGNs do
  that or not, but they definitely have the means. It is also possible
  that the AGN outflows get stopped in the central parts of galaxies,
  never reaching cosmological scales, or that, even if they reach
  cosmological scales, outflows expand more quickly into low density
  regions without affecting higher density regions much.

  Additionally, AGNs clustering might influence the power on large
  scales.  Even if AGNs cause non-gravitational fluctuations only on
  smaller, sub-Mpc scales, these fluctuations propagate to larger,
  tens of Mpc scales, because the AGNs themselves are clustered more
  than the baryons as a whole.
                                                                                
  Admittedly, our model of spherical outflows is excessively
  simplistic, and should probably only be considered as an upper
  limit. Nevertheless, it presents a counterexample to the widespread
  belief that lensing measurements are insensitive to complex
  astrophysics.
                                                                                           
  Unfortunately, the interplay between the two competing effects of
  AGNs is quite intricate, so even the sign of the total effect of AGN
  outflows on the matter power spectrum cannot be deduced without
  further detailed numerical studies of complex gas dynamics on
  galactic and sub-galactic scales. But the final success of future
  weak lensing studies of the dark energy will substantially depend on
  our ability to make a theoretical breakthrough in modeling AGN
  outflows on a wide range of scales.
 

  \bibliography{mpowerpap}

  
\end{document}